\documentclass{jltp} % -*- LaTeX -*-
\usepackage{graphicx}
\usepackage{calc}
\usepackage{amsmath}
\usepackage{amssymb}
\usepackage{bm}

\begin{document}

\def\topfraction{1} \def\textfraction{0}

\title{Vortex formation in neutron-irradiated rotating superfluid $^3$He-B}

\author{A.P.~Finne$^{*}$, S.~Boldarev$^{*,\dagger}$, V.B.~Eltsov$^{*,\dagger}$, and 
M.~Krusius$^*$}

\address{$^*$Low Temperature Laboratory, Helsinki University of
  Technology,\\ P.O.
  Box 2200, FIN-02015 HUT, Finland\\
  $^\dagger$Kapitza Institute for Physical Problems, 119334 Moscow, Russia}
\vspace{-6mm}

\runninghead{A.P.~Finne \textit{et al.}}{Vortex formation in neutron irradiation}

\maketitle \vspace{-6mm} \begin{abstract}
  A convenient method to create vortices in meta-stable vortex-free
  superflow of $^{\mathit{3}}\!$He-B is to irradiate with thermal
  neutrons. The vortices are then formed in a rapid non-equilibrium process
  with very distinctive characteristics. Two models were suggested to
  explain the phenomenon. One is based on the Kibble-Zurek mechanism of
  defect formation in a quench-cooled second order phase transition.
  The second model builds on the instability of the moving front between
  superfluid and normal $^{\mathit{3}}\!$He, which is created by the 
  heating from the neutron absorption event. The most detailed measurements
  with single-vortex resolution have been performed at temperatures close to $T_c$. 
  We present an overview of the main experimental features and demonstrate that 
  the measurements are consistent with the Kibble-Zurek picture. New data, 
  collected at low temperatures, support
  this conclusion, but display superfluid turbulence as a new phenomenon.
  Below $\mathit{0.6}\,T_c$ the damping of vortex motion from the normal
  component is reduced sufficiently so that turbulent vortex dynamics become
  possible. Here a single absorbed neutron may transfer the sample from the
  meta-stable vortex-free to the equilibrium vortex state. We find that the
  probability for a neutron to initiate such a turbulent transition grows
  with increasing superflow velocity and decreasing temperature.

PACS numbers: 47.32, 67.40, 67.57, 98.80\\ %Keywords: quantized vortex; 
%non-equilibrium phase transition; superfluid turbulence; critical velocity 
\end{abstract} \vspace{-12mm}

\section{NON-EQUILIBRIUM PHASE TRANSITIONS}

Exposure to radiation causes changes in the structure of matter --
radiation damage -- which depends on the type of radiation and the absorbing
material. Defects are produced, which may range from nuclear reactions to
ionization and other non-elastic scattering effects. The interaction
products then have to be accommodated in the structure of the material
which leads to a multitude of different phenomena. In superconductors and
superfluids absorption events from applied radiation induce defects also in
the spatial distribution of the order parameter. If the state of such a
coherent many-body system is close to a critical point, for instance in a
superfluid in the regime where the flow may intermittently switch between
laminar and turbulent, the life time of the laminar state has been found to
be limited by the background radiation level of the laboratory
surroundings.\cite{Schoepe} We discuss here a similar case where applied
ionizing radiation is found to create quantized vortex lines in meta-stable
superflow which is originally prepared to be vortex-free.\cite{Ruutu}
Depending on conditions, one irradiation event may produce from zero to a
few vortex lines, or it may suddenly send the sample to the equilibrium 
vortex state, producing thousands of vortices.

An ionization event leads to heating where some small volume of the superfluid is 
abruptly heated above $T_{\rm c}$ to the normal state. The subsequent cool down of 
this warm bubble back to the superfluid state is an example of a rapid phase 
transition in conditions far from equilibrium. In 1976 Tom Kibble proposed that the 
inhomogeneous distribution of visible matter in the universe -- known as the 
large-scale structure -- might originate from defects which are formed in a rapid 
2nd order phase transition during the early expansion and cooling after the Big 
Bang.\cite{Kibble} The defect formation transforms the original homogeneous system 
into an inhomogeneous. This suggestion was augmented by Wojciech Zurek in 1985 with 
quantitative predictions for the density of the defects, using scaling arguments 
about the slowing down of the order-parameter relaxation at the critical 
point.\cite{Zurek} Since then, cosmic strings have not been found and measurements 
on the angular distribution of the anisotropy in the cosmic background radiation 
have supported another explanation -- the inflation model -- as the origin of 
large-scale structure.  

In condensed matter physics the Kibble-Zurek (KZ) mechanism of defect formation has 
remained an important issue: Are there real laboratory examples in which this 
mechanism can be proven to work?  Phase transitions in condensed-matter systems are 
more often than not connected with defect formation. This is ascribed to different 
inhomogeneities of the system, such as impurities, grain boundaries, surfaces, etc. 
Superfluid $^3$He-B was perhaps the first system where the KZ mechanism has been 
difficult to dispute,\cite{Ruutu,Bunkov,Eltsov,SpinMass} the match between 
experiment and model looks perfect. However, solid theoretical justification is 
still missing which would prove the process in superfluid $^3$He-B with a detailed 
microscopic calculation.

Ions in superfluid $^4$He-II can be accelerated with electric fields to produce 
vortex rings.\cite{RayfieldReif} The ions formed in applied radiation can thus be 
assumed to be directly responsible for the vortex production if electric fields are 
present. This is assumed to be the case in Ref.~\onlinecite{Schoepe}. In zero 
applied electric field the experimental situation in $^4$He-II is 
controversial,\cite{McClintock} while in $^3$He-B irradiation with neutrons at 
thermal wave lengths has become a useful tool for creating vortex lines in 
vortex-free superflow. 

In liquid $^3$He the mean length of flight of a thermal neutron is $\sim 0.1\,$mm, 
before the neutron undergoes the capture reaction n + $^3_2$He $\rightarrow$ p + 
$^3_1$H + 764\,KeV. The reaction products, the proton and the triton, carry the 
kinetic energy of 764\,keV and produce a cascade of ionizations among the 
surrounding $^3$He atoms. The ionization processes and the resulting heating are 
restricted to a volume of less than 0.1\,mm in diameter. The heated bubble cools 
back to the temperature of the surrounding superfluid bath within microseconds. 
Thus the heating occurs locally within the bulk superfluid, but close to the outer 
boundary of the cylindrical sample (with radius $R$).  In rotation at the angular 
velocity $\Omega$ this is the place with a well-defined superflow velocity, $v_{\rm 
s} = \Omega R$, the largest possible value. If this superflow velocity exceeds some 
critical value, then vortices are observed to emerge from a single neutron 
absorption event. In practice the process has turned out to be a conveniently 
controllable method to form vortices.  Its critical velocity is unaffected by the 
surface properties of the sample container. If other critical velocities happen to 
be unattainably high, neutron radiation can always be used to create vortices in a 
predictable way. Experimentally a major advantage of neutrons over other types of 
radiation, such as $\gamma$ rays, is that temperature stability can be maintained 
without any effort, since neutron absorption elsewhere in the apparatus is 
insignificant.

This report is divided in two main sections. The first contains an analytic model 
of the KZ mechanism which is then compared to experiment in the high temperature 
regime $(0.80\,T_{\rm c} < T < T_{\rm c})$ where the measurements have been most 
extensive.\cite{Ruutu,PLTP-chapter} The second section reports on new experiments 
below $0.60\,T_{\rm c}$. Our measurements are performed with NMR techniques. In the 
high temperature range the vortex line count can be carried out with single-vortex 
resolution. The vortex dynamics in this temperature range is severely damped by 
mutual friction and is reasonably well understood. Thus it is possible to trace the 
connection between the initial configuration of vortices after the neutron 
absorption event and their final number as rectilinear lines in rotation. Here the 
agreement between the KZ model and the experiment is at least semi-quantitative.  
In contrast, below $0.60\,T_{\rm c}$ the injection of the vorticity by the neutron 
absorption event sends the $^3$He-B sample with some probability for a short 
transitory period into a turbulent state, where the number of vortices rapidly 
multiplies. After the relaxation of the turbulence, the final stable state is the 
equilibrium vortex state. In this process the initial configuration of vortices 
after the neutron absorption event is sufficient to trigger the turbulence and is 
thus not directly related to the final number of rectilinear lines.

\section{MODEL OF VORTEX FORMATION} \label{AnalyticModel}

\subsection{Defect formation in a non-equilibrium phase transition}
\label{UniformQuench}

A homogeneous thermal quench through the transition temperature $T_{\rm c}$ is
characterized in the KZ model\cite{Zurek} by one experimental variable, the quench time
\begin{equation}
        \tau_{\rm Q} = \left( \frac{1}{T_{\rm c}}
          \left|{\frac{dT}{dt}}\right|_{T=T_{\rm c}} \right)^{-1}~~,
\label{QuenchTime}
\end{equation}
so that the temperature evolution $T(t)$ at $T_{\rm c}$ can be approximated with a linear
dependence $T = T_{\rm c} (1 - t/\tau_{\rm Q})$. The quench time $\tau_{\rm Q}$ has to be
compared to the order parameter relaxation time $\tau (T)$, which for a Ginzburg-Landau
system at a second order phase transition is assumed to be of the form
\begin{equation}
     \tau (T) = \tau_0 (1 - T/T_{\rm c})^{-1}~~.
\label{RelaxTime}
\end{equation}
In superfluid $^3$He, $\tau_0$ is on the order of $\tau_0 \sim \xi_0/v_{\rm F}$. Here
$\xi_0$ is the zero temperature limiting value of the temperature $(T)$ and pressure
$(P)$ dependent superfluid coherence length $\xi(T,P)$, while $v_{\rm F}$ is the velocity
of the thermal quasiparticle excitations which are excited above the superfluid energy
gap. Close to $T_{\rm c}$ in the Ginzburg-Landau temperature regime, we have $\xi(T,P) =
\xi_0(P) (1-T/T_{\rm c})^{-1/2}$. Thus below $T_{\rm c}$ the order parameter coherence
can be assumed to spread out with the velocity $c(T) \sim \xi/\tau = \xi_0 \, (1-T/T_{\rm
c})^{1/2}/ \tau_0$. The freeze-out of defects occurs at time $t_{\rm Z}$, when the
causally disconnected regions have grown together and superfluid
coherence becomes established in the whole volume. At the corresponding
freeze-out temperature $T_{\rm Z} = T(t_{\rm Z}) < T_{\rm c}$, the causal
horizon has travelled the distance $\xi_{\rm H} (t_{\rm Z}) =
{\int}_0^{t_{\rm Z}} \, c(T) \, dt = \xi_0 \tau_{\rm Q}
(1-T_{\rm Z}/T_{\rm c})^{3/2}\, / \tau_0$ which has to be
equal to the coherence length $\xi(t_{\rm Z})$. This condition
establishes the freeze-out temperature $T_{\rm Z}/T_{\rm c} =
1 - \sqrt{\tau_0/\tau_{\rm Q}} \;$ at the freeze-out time $t_{\rm
Z} = \sqrt{\tau_0 \tau_{\rm Q}}\;$, when the domain size has
reached the value
\begin{equation}
  \xi_{\rm v} = \xi_{\rm H}(t_{\rm Z}) =
  \xi_0 \; (\tau_{\rm Q}/\tau_0)^{1/4}~.
\label{eq:xi-initial}
\end{equation}
Thus the causal horizon, behind which superfluid coherence is established,
travels with the velocity
\begin{equation}
  v_{\rm Tc} = {\frac{\xi_{\rm H}(t_{\rm Z})} {t_{\rm Z}} }=
 {\frac{ \xi_0} {\tau_0}} \;
\left({\tau_0 \over \tau_{\rm Q}}\right)^{1/4}~.
\label{eq:Tc-velocity}
\end{equation}

Characteristic numbers for superfluid $^3$He are $\xi_0 \sim 20$ nm,
$\tau_0 \sim 1$~ns, and a quench time of $\tau_{\rm Q} \sim 1$ $\mu$s. From
these values we expect the domain structure to display a characteristic
length scale of order $\xi_{\rm v} \sim 0.1$ $\mu$m. In a
U(1)-symmetry-breaking transition, vortex lines are expected to form at the
domain boundaries. This leads to a network of randomly organized vortices,
where the average inter-vortex distance and radius of curvature are on the
order of the domain size $\xi_{\rm v}$. In general, the KZ scaling model
predicts a rapid homogeneous quench to produce a defect density (defined as
vortex length per unit volume) $l_{\rm v} = ({a_l \xi_{\rm v}^2})^{-1}.$
The numerical factor $a_l \sim 1$\,--\,100 depends on the details of the
model system.

\subsection{Threshold velocity for vortex loop escape} \label{LoopEscape}

The number of vortex loops which can be extracted from a heated neutron bubble 
depends on the applied superflow velocity $v_{\rm s}$. More precisely, in the 
rotating experiments the applied flow is the counterflow velocity ${\bf v} = {\bf 
v}_{\rm s} - {\bf v}_{\rm n}$, where the flow velocity of the viscous normal 
component is ${\bf v}_{\rm n} = {\mathbf \Omega} \times {\bf r}$, when the $^3$He 
sample of radius $R$ rotates at the angular velocity $\Omega$. In fact, the neutron 
capture events are monitored at constant rotation $\Omega$, so that the normal 
component is firmly clamped to corotate with the container. In the rotating frame 
of reference we then have $v_{\rm n} =0$ and $v = v_{\rm s}$. The counterflow 
velocity $v$ is the applied bias which determines the number of vortex loops to be 
extracted from the heated neutron bubble. In particular, there exists a threshold 
value $v_{\rm cn}$ for the bias, below which it will not suffice to pull vortex 
loops from the bubble. The dependence on the rotating counterflow bias $v$ can be 
studied from the threshold $v_{\rm cn}$ up to the critical limit $v_{\rm c}$ at 
which an elemental vortex loop is spontaneously formed at the cylindrical wall in 
the absence of the neutron flux.\cite{Parts}

Thus the threshold $v_{\rm cn}$ is the smallest bias velocity at
which a vortex ring can escape from the heated neutron bubble. It can be
connected with the size of the bubble in the following manner. A vortex
ring of radius $r_\circ$ is in equilibrium in the applied counterflow at
the velocity $v$ if it satisfies the equation
\begin{equation}
  r_\circ(v) = {\kappa \over {4\pi v}} \; \ln{\left(
    {r_\circ \over \xi(T,P)} \right)} \, ,
\label{VorRing}
\end{equation}
where $\kappa = 6.61\cdot 10^{-4}\,\textrm{cm}^2/\textrm{s}$ is the
circulation quantum.  This follows from the balance between self-induced
contraction and expansion by the Magnus force: a ring with a radius larger
than $r_\circ$ will expand in the flow while a smaller one will contract.
Thus the threshold or minimum velocity at which a vortex ring can start to
expand towards a rectilinear vortex line corresponds to the maximum
possible vortex-ring size. This must be of a size comparable to the
diameter of the heated bubble. For a simple estimate we set the vortex ring
radius equal to that of a spherical neutron bubble: $r_\circ (v_{\rm cn})
\sim R_{\rm b}$, thus $v_{\rm cn} \propto 1/R_{\rm b}$.

A simple thermal diffusion model can be used to estimate the magnitude of the 
radius $R_{\rm b}$ of the bubble which originally was heated above $T_{\rm c}$. In 
the temperature range close to $T_{\rm c}$ the final phase of cooling occurs via 
diffusion of quasiparticle excitations out into the surrounding superfluid bath 
with a diffusion constant $D \approx v_{\rm F} l$, where $v_{\rm F}$ is their Fermi 
velocity and $l$ their mean free path. For a spherically symmetric temperature 
profile $T(r,t)$ as a function of the radial distance $r$ and time $t$, the 
diffusion equation is \begin{equation} {{\partial T(r,t)} \over
  {\partial t}} = D \left( {{\partial^2 T} \over {\partial r^2}} + {2 \over
  r} {{\partial T} \over {\partial r}} \right) \; .
\label{TherDiff}
\end{equation}
With the assumption that at $t=0$ the energy $E_0$ is deposited at
$r=0$, the solution is given by
\begin{equation}
  T(r,t) - T_0 \approx {E_0\over C_{\rm v}} \;
  {1\over (4 \pi D t )^{3/2}} \; \exp \Biggl ({-r^2\over 4Dt} \Biggr ),
\label{e.1} \end{equation} where $T_0$ is the temperature of the surrounding 
superfluid bath and $C_{\rm v}$ is the specific heat. The energy $E_0$ which is 
deposited as heat is close to the energy released in the nuclear reaction. (It is 
believed that $\sim 10$\,\% of the reaction energy is turned into ultraviolet 
radiation and retarded relaxation of excited molecular complexes, ie.  into 
components which do not contribute to the heating of the neutron bubble.) The 
bubble of normal fluid, $T(r) > T_{\rm c}$, expands and reaches a maximum radius 
\begin{equation} R_{\rm b} = \sqrt {3 \over {2 \pi e}} \; \left( {E_0 \over {C_{\rm 
v} T_{\rm c}}} \right)^{1/3} \;  (1-T_0/T_{\rm c})^{-1/3} \; . \label{e.2} 
\end{equation} It then starts cooling and rapidly shrinks with the characteristic 
time $\tau_{\rm Q} \sim R_{\rm b}^2/D \sim 1 \mu$s. Since $v_{\rm cn}$ is inversely 
proportional to $R_{\rm b}$, it has the temperature dependence $v_{\rm cn} \propto 
(1-T_{0}/T_c)^{1/3}$ close to $T_{\rm c}$. Also, the prefactor of Eq.~(\ref{e.2}) 
decreases with increasing pressure as both $C_{\rm v}$ and $T_{\rm c}$ grow. Thus 
$v_{\rm cn}$ should increase with pressure.  These predictions for the temperature 
and pressure dependences of $v_{\rm cn}$ agree semi-quantitatively with 
measurements.\cite{Ruutu,Eltsov}

\subsection{Escape rate of vortex loops} \label{AnalVorEsc}

During the quench-cooling of the neutron bubble through the superfluid transition, 
a random vortex network is formed within the bubble.\cite{Zurek} According to the 
KZ model,\cite{Zurek} the characteristic length scale of the network (i.e. the 
average inter-vortex distance and average radius of curvatures of the vortices) is 
comparable to the order parameter inhomogeneity $\xi_{\rm v}$, which is 
precipitated in the quench. The later evolution of the network leads to a gradual 
increase in this length. We shall call this time-dependent length $\tilde \xi(t)$. 
In the following we assume that this ``coarse-graining'' process of the network 
preserves its random character, in other words the network remains self similar or 
scale invariant. Only later a change occurs in this respect, when the loops become 
sufficiently large to interact with the externally applied bias field: This causes 
large loops with a radius exceeding the critical value from Eq.~(\ref{VorRing}) to 
expand, if they are oriented transverse to the flow with the correct winding 
direction.  Eventually such loops are extracted to the bulk and grow to become 
rectilinear vortex lines in the center of the sample. (At high temperatures in the 
absence of turbulence the number of vortices is conserved, after the evolution of 
the random network is finished and well-defined loops in the bulk bias field have 
been formed.) The small loops in the network contract and disappear. The time scale 
of these processes is determined by the mutual friction damping between the normal 
and superfluid components. In the high temperature range the evolution of the 
vortex network occurs in milliseconds and the growth of the extracted vortex rings 
into rectilinear lines happens in a fraction of a second.

The number of vortex loops, which are extracted from the network and are observed 
in the measurement, can be found from the following considerations.  The energy of 
a vortex loop, which is stationary with respect to the walls, is given 
by\cite{Donnelly} 
\begin{equation} {\cal E}=E_{\rm kin}+{\bf p} {\bf v}\,, 
\label{eq10} \end{equation} 
where ${\bf v}$ is the velocity of the bias flow. The 
hydrodynamic kinetic energy or self-energy of the loop arises from the trapped 
superfluid circulation with the velocity $v_{\rm s,vort}$, \begin{equation} E_{\rm 
kin}={1\over 2}\int \rho _{\rm s} v^2_{\rm s,vort} \, dV= \varepsilon L~, 
\label{eq11} \end{equation} and is proportional to the length $L$ of the loop and 
its line tension, \begin{equation} \varepsilon  = {{\rho_{\rm s} \kappa^2} \over 
{4\pi}} \; \ln
 {\tilde\xi(t) \over {\xi}}~~.\label{eq12}
\end{equation}
Here we neglect the small contribution from the core energy, use
$\tilde\xi(t)$ for the diameter of the loop, and the superfluid coherence
length $\xi(T,P)$ for the diameter of the core.  This equation is valid in
the logarithmic approximation, when $\tilde\xi(t) \gg \xi(T,P)$.  While the
first term in Eq.~(\ref{eq10}) is proportional to the length $L$ of the
loop, the second term involves its linear momentum,
\begin{equation}{\bf p} = \int \rho_{\rm s} {\bf v}_{\rm s,vort} \,dV= {1
\over 2\pi} \, \rho_{\rm s} \kappa \,  \int {\bf \nabla} \Phi\, dV= \rho_{\rm s}
\kappa {\bf S}~~,\label{eq13}
\end{equation}
where the last step follows from Gauss's theorem and involves the area $S$
of the loop in the direction of the normal ${\bf S}/S$ to the plane of the
loop. Thus we write for the energy of a loop
\begin{equation}
  {\cal E}(L,S,t) =\rho_{\rm s} \kappa ~\left[ ~L~{\kappa \over 4\pi} \; \ln
 {\tilde\xi(t) \over {\xi}}~-~ v S ~\right]~~,
\label{eq14}
\end{equation}
where $S$ is now the algebraic area perpendicular to the flow and of proper
winding direction. This equation expresses the balance between a
contracting loop due to its own line tension, which dominates at small bias
velocities, and expansion by the Magnus force from the superflow, which
dominates at high bias velocities. The divide is the equilibrium condition,
which was expressed by Eq.~(\ref{VorRing}) and corresponds to the situation
when the height of the energy barrier, which resists loop expansion,
vanishes. In this configuration ${\bf p}$ is antiparallel to ${\bf
  v}$, the loop moves with the velocity $-{\bf v}$ in a frame of the
superfluid component, but is stationary in the rotating frame.

The expansion of the vortex loop should be calculated by including the
mutual friction forces. In our analytic description of vortex loop escape
we shall neglect such complexity. Instead we shall make use of three
scaling relations which apply to Brownian networks\cite{Vachaspati} and are
derived from simulation calculations described in
Ref.~\onlinecite{PLTP-chapter}. These expressions relate the mean values in
the statistical distributions of the loop diameter ${\cal D}$, area $S$,
and density $n$ to the length $L$ of the loop:
\begin{equation}
  {\cal D} = A L^\delta\,\tilde\xi^{1-\delta}~,
  \quad (A \approx 0.93,\;\;\delta \approx
  0.47)~,
\label{DLx} \end{equation}
\begin{equation}
  |S| = B {\cal D}^{2-\zeta}\,\tilde\xi^\zeta,
  \quad (B \approx 0.14,\;\;\zeta \approx 0)~,
\label{SDx} \end{equation}
\begin{equation}
  n = CL^{-\beta}\,\tilde\xi^{\beta-3}~,
  \quad (C \approx 0.29,\;\;\beta \approx 2.3)~.
\label{NLx}
\end{equation}
For a Brownian random walk in infinite space the values of $\delta$, $\beta$ and
$\zeta$ are $1/2$, $5/2$ and 0. The important assumption is that these relations are valid
during the entire evolution of the network, until sufficiently large rings are extracted
by the counterflow into the bulk. Using Eqs.~(\ref{DLx}) and (\ref{SDx}), we may write
Eq.~(\ref{eq14}) for the energy of a loop in the form
\begin{equation}
  {\cal E} ({\cal D},t) =
  \rho_{\rm s} \kappa {\cal D}^2~\left[ ~{\kappa\over 4\pi
      \tilde\xi(t)A^2} \; \ln {\tilde\xi(t) \over {\xi}}~-~
    v B ~\right]~~.
\label{eq15} \end{equation} When the mean diameter $\tilde \xi(t) $ exceeds a 
critical size $\tilde \xi_{\rm c} (v)$, which depends on the particular value of 
the bias velocity $v$, \begin{equation} \tilde \xi_{\rm c} (v)= { 1 \over {A^2B}} 
\; {\kappa \over {4\pi v}} \; \ln {\tilde \xi_{\rm c} \over {\xi}}~~~,\label{eq16} 
\end{equation} the energy in Eq.~(\ref{eq15}) becomes negative and the loop starts 
expanding spontaneously. This is the smallest loop which will be able to expand at 
a given value $v$. The upper cutoff for the loop size distribution is provided by 
the diameter of the entire network, or that of the heated bubble, $2R_{\rm b}$, 
such that $\tilde\xi_{\rm c}(v_{\rm cn}) = 2 R_{\rm b}$. The total number of loops 
$N_{\rm b}$, which will be extracted from one neutron bubble, can then be obtained 
from \begin{equation} N_{\rm b} = V_{\rm b} \, \int_{\tilde\xi_{\rm c}}^{2R_{\rm 
b}}d{\cal D}~n({\cal D})~. \label{eq17} \end{equation} Here the density 
distribution $n(L)=C~\tilde \xi^{-3/2}~L^{-5/2}$, combined with that for the 
average diameter ${\cal D}(L) =A~(L~ \tilde \xi ~)^{1/2}$, gives $n({\cal D}) \, 
d{\cal D} = 2A^3C{\cal D}^{-4} \; d{\cal D}$. On inserting this into the 
integral~(\ref{eq17}) we obtain \begin{equation} N_{\rm b} = {1 \over 9} \pi A^3 C 
\left[\left({{2R_{\rm b}} \over \tilde\xi_{\rm c}}\right)^3 -1\right] \; . 
\label{eq18} \end{equation} From this equation we see that the requirement $ N_{\rm 
b}(v_{\rm cn})=0$ returns us the definition of the threshold velocity $ v_{\rm 
cn}$: $\tilde \xi_{\rm c} (v=v_{\rm cn}) = 2 R_{\rm b}$. This in turn gives us from 
Eq.~(\ref{eq16}) for the radius of the heated bubble \begin{equation} R_{\rm b} = { 
1 \over {A^2B}} \; {\kappa \over {8\pi v_{\rm cn}}} \; \ln {2R_{\rm b} \over 
{\xi(T,P)}}~~~\label{eq19} \end{equation} Eqs.~(\ref{eq16}) and (\ref{eq19}) show 
that $\tilde \xi_{\rm c} \propto 1/v$ and $R_{\rm b} \propto 1/ v_{\rm cn}$, so 
that we may write for the vortex-formation rate $\dot N =\phi_{\rm n} N_{\rm b}$ 
from Eq.~(\ref{eq18}) \begin{equation} {\dot N} = {1 \over 9} \pi A^3 C \phi_{\rm 
n} \, \left[\left({v \over v_{\rm cn}}\right)^3 -1\right] \; , \label{eq20} 
\end{equation} where $\phi_{\rm n}$ is the neutron flux. Thus the rate, at which 
vortex loops are extracted into the bias flow under neutron irradiation, has a cubic 
dependence on the bias velocity $v$ and reflects the dependence on the volume of the 
heated neutron bubble. By inserting $A \approx 0.93$, $C \approx 0.29$ from 
Eqs.~(\ref{DLx}) and (\ref{NLx}), respectively, and $\phi_{\rm n} \approx 20$ 
neutrons/min (as appropriate for the saturation of the event rate $\dot N_e$  in the 
center panel of Fig.~\ref{BiasDependence}), we obtain for the prefactor in 
Eq.~(\ref{eq20}) $\gamma = {1 \over 9} \pi A^3 C \phi_{\rm n} \approx 1.6$ 
min$^{-1}$. We may also define a threshold velocity $v_{\rm cni}$ for an event in 
which $i$ loops are formed simultaneously, using the approximate requirement 
$N_{\rm b} (v=v_{\rm cni}) \approx i$. This gives $v_{{\rm cn}i}/v_{\rm cn}\sim 
i^{1/3}$. These results are found to be consistent with measurements.

\subsection{Non-uniform thermally driven transition} \label{NonuniformQuench}

The cubic dependence on the bias flow in Eq.~(\ref{eq20}) comes from the assumption that
the whole volume of the heated bubble contributes equally to the production of vortex
loops. In the rapidly cooling neutron bubble there is a steep thermal gradient such that
the causal horizon, behind which superfluid coherence is established, lags behind
the temperature front, where $T$ drops below $T_{\rm c}$. To characterize such a
non-uniform quench,\cite{Volovik} we need in addition to the quench time $\tau_{\rm Q}$ a
second parameter, the thermal length scale $\lambda = [|\nabla T| / T_{\rm c}]^{-1}$. The
temperature front can then be assigned a velocity of order $v_{\rm T} \sim
\lambda/\tau_{\rm Q}$. If the transition is slow, the
causal horizon will keep abreast of the thermal front and superfluid coherence
will not be broken. However, in a rapid non-equilibrium situation $v_{\rm T} > v_{\rm
Tc}$ and the  KZ mechanism will survive in the space between the two separated fronts.

In the case of the heated neutron bubble $\lambda \sim R_{\rm b} \sim 50 \,\mu$m and so
$v_{\rm T} \sim 50\,$m/s, which is of the same order of magnitude as $v_{\rm
Tc}$ from Eq.~(\ref{eq:Tc-velocity}). Thus we do not expect a serious suppression in
vortex formation owing to the nonuniformity of the temperature distribution.

\subsection{Alternative explanations of vortex formation}
\label{Volume/Surface}

The neutron-capture experiment in rotating $^3$He-B is not an exact replica of the 
ideal KZ model, a quench-cooled second order transition in an infinite homogeneous 
medium. In the neutron bubble there is a strong thermal gradient and a strict 
boundary condition applies at its exterior, imposed by the bulk superfluid state 
outside. The cool-down occurs so fast that any extrapolation from the equilibrium 
state theories is uncertain, whether it concerns the hydrodynamics or the 
superfluid state.  Non-equilibrium phase transitions are notoriously a complicated 
issue and any interpretation has to be questioned. Nevertheless, the predictions of 
the KZ model, as summarized above, are semi-quantitatively supported by the 
experimental observations.\cite{PLTP-chapter} One may still argue whether this fact 
constitutes final proof or not: Perhaps other processes can be found which also 
explain the experimental observations?

Besides the KZ mechanism, the only alternative model which at the moment exists on 
the level of quantitative predictions is a surface instability -- the production of 
vortex rings around the circumference of the heated neutron bubble, at the 
interface between the surrounding cold superfluid bath and the hot normal liquid in 
the center. This model was worked out by Aranson {\it et al.}  
(Ref.~\onlinecite{Kopnin}). They demonstrated that the moving normal-superfluid 
interface becomes unstable in the presence of the superflow along the interface. 
Numerical simulations show that the outcome of this instability is the production 
of vortex rings which encircle the bubble and are perpendicular to the applied 
flow. These rings screen the superflow, so that the superfluid velocity (in the 
rotating frame) is zero inside the bubble.

It is obvious that the possible influence of the surface instability has to be 
examined carefully: During the cooling the fluid in the shell around the normal 
bubble still remains in the B phase but is heated above the surrounding bulk 
temperature $T_0$. In the Ginzburg-Landau temperature regime the intrinsic 
instability velocity $v_{\rm cb}(T,P)$ of the bulk superfluid decreases with 
increasing temperature.\cite{Parts} If no other process intervenes, the superflow 
instability therefore necessarily has to occur within that peripheral warm shell 
surrounding the hot neutron bubble, where the local temperature corresponds to the 
temperature at which the bias velocity $v$ equals the critical value $v_{\rm 
cb}(T,P)$ for bulk superflow.

After their formation, the rings, which are produced by the surface instability, 
start to expand and will eventually be pulled away by the Magnus force. It is thus 
conceivable that these rings might give rise to the rectilinear vortex lines which 
are observed in the NMR measurements. How do we distinguish from the measured data 
whether the surface instability or the KZ mechanism is responsible for the observed 
vortex lines? The critical velocity $v_{\rm cn}$ does not discriminate between the 
two processes. In both cases it is determined by the stability velocity 
(Eq.~(\ref{VorRing})) for the largest vortex ring which fits the bubble. However, 
the different nature of the two processes has important consequences for the vortex 
formation rate.

The row of vortex rings, which is produced by the surface instability to screen the 
neutron bubble, resembles a vortex sheet between the superfluid, moving with 
velocity $v$ outside the bubble, and the stationary superfluid inside. The density 
of vorticity in the sheet is $v/\kappa$ and the number of loops produced by one 
neutron absorption event is $N_{\rm b} \propto v R_{\rm b}/\kappa$, where $R_{\rm 
b}$ is the size of the bubble along the flow direction. Thus $N_{\rm b}$ grows 
linearly with the applied flow velocity. This estimate is supported by the 
numerical simulations.\cite{Kopnin} In contrast, the KZ mechanism is a volume 
effect, which results in the cubic dependence expressed in Eq.~(\ref{eq20}). This 
is a major distinguishing feature between the two mechanisms, from the experimental 
point of view.

In addition, the surface instability should be a deterministic process which in 
every event produces the same surface density of vortex rings.  Variations in the 
number of rings arises only owing to variations in the shape of the neutron bubble 
and its orientation with respect to the bias flow. (Unfortunately, the distribution 
of the number of rings was not studied in detail in the simulations.\cite{Kopnin}) 
In contrast, the KZ mechanism produces a random vortex network. The number of loops 
which is extracted by the applied flow from such a  network is inherently a 
stochastic number with a relatively wide distribution.

Finally, the surface instability is not particularly sensitive to the processes 
inside the heated neutron bubble. Only regular mass-flow vortices are expected to 
form. The KZ mechanism, on the other hand, can be expected to produce all possible 
different kinds of defects. Their presence might be either directly observed in the 
final state or via their influence on the evolution of the vortex network inside 
the neutron bubble. 

All these features can be checked in the experiment. As outlined in the next 
paragraph, the experimental results are undoubtedly more consistent with the KZ 
mechanism. This fact is puzzling: From the two processes the surface instability 
should be the dominant one. As long as the rings created by the surface instability 
encircle the neutron bubble, they shield its central volume from the bias flow and 
any vorticity within the bubble -- which might have been created by the KZ 
mechanism -- will collapse in the presence of dissipation. This is exactly what was 
seen in the numerical simulations\cite{Kopnin} with a rapidly cooling model system 
represented by a scalar order parameter. The thermal diffusion equation was used to 
account for the cooling and the order-parameter relaxation was obtained from the
time-dependent Ginzburg-Landau equation. It was then found that both processes give 
rise to vortex formation, but the rings, which eventually manage to escape into the 
bulk, originate from the bubble boundary.  

The discrepancy between simulation and experiment demonstrates that the competition 
between the two vortex formation mechanisms is not as straightforward as described 
above. Possibly in the conditions of the experiment the surface instability 
develops so slowly that the random KZ network manages to form earlier. The 
polarization of the network by the bias flow might then stop the development of the 
surface instability. Also the diffusion model, which is used both here in 
Sec.~\ref{LoopEscape} and in Ref.~\onlinecite{Kopnin} for the cooling of the 
bubble, is a gross  oversimplification.\cite{Leggett} This fact might be another  
important source for the discrepancy. 

\subsection{Comparison with experiment}  \label{KZ-evidence}

%%%%%%%%%%%%%%%%%%%%%%%%%%%%%%%%%%%%%%%%%%%%%%%%%%%%%%%%%%%%%%%%%%%%%%%%%%%%%%%
\begin{figure}[!!!tb]
  \centerline{\includegraphics[width=1.0\columnwidth]{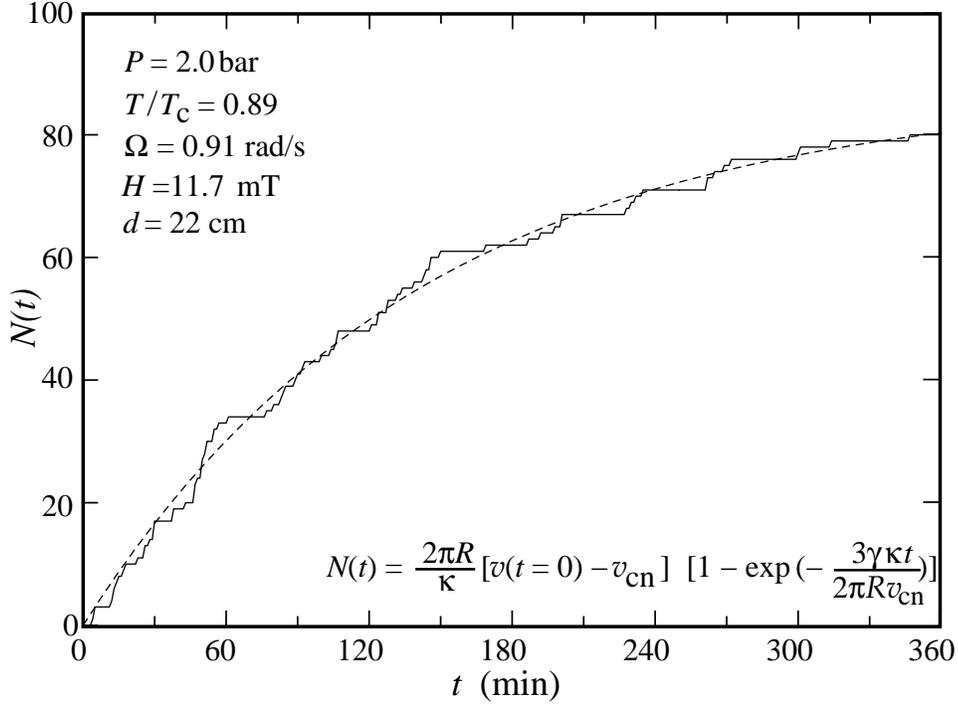}}
  \bigskip
\caption[Saturation] {
  Measurement of vortex formation in neutron irradiation at high
  temperatures. Each discontinuous step in the measured staircase trace
  corresponds to a different neutron absorption event. The height of the
  step gives the yield of rectilinear vortex lines from this event. At
  $t=0$ the neutron irradiation is turned on at constant flux on the
  initially vortex-free $^3$He-B sample rotating at constant $\Omega$. The
  dashed curve represents the expression shown in the panel which
  has been integrated from the rate Eq.~(\protect\ref{eq20}). The values of 
  the two parameters, $\gamma = 1.1$ min$^{-1}$ and $v_{\rm cn} = 1.9$ mm/s, 
  have been chosen to give a good fit, but agree within the measuring 
  accuracies with all other measurements in the temperature range from 
  $T_{\rm c}$ down to $0.80\,T_{\rm c}$. The data in Figs.~\ref{BiasDependence} 
  and \ref{VorRingCount} have been collected by measuring the initial slope 
  of this curve, ${\dot N}(t)$ at $t=0$. }
   \label{Saturation}
  %\end{center}
\vspace{-6mm}
\end{figure}
%%%%%%%%%%%%%%%%%%%%%%%%%%%%%%%%%%%%%%%%%%%%%%%%%%%%%%%%%%%%%%%%%%%%%%%%%%%

The dependence of the formation rate ${\dot N}$ of rectilinear vortex lines on the 
externally applied bias velocity $v = |\mathbf{v}_{\rm s} - \mathbf{v}_{\rm 
n}|_{r=R} = \Omega R - \kappa N/(2\pi R)$ is the central quantitative result from 
the rotating measurements. It is most efficiently measured close to $T_{\rm c}$ 
where the resolution in NMR absorption is sufficient to resolve individually the 
rectilinear vortex lines which are formed in each neutron capture event.  

Fig.~\ref{Saturation} illustrates a measurement where the cumulative number of 
vortex lines $N(t)$ has been recorded as a function of time $t$ over a period of 6 
hours.\cite{LT22} Initially, when the neutron flux is turned on, the sample is 
rotating at constant angular velocity in the vortex-free state.  First the vortex 
formation rate ${\dot N}$ is high and individual neutron absorption events produce 
several rectilinear vortex lines. The lines accumulate in a dense cluster which 
lies coaxially in the center of the cylindrical sample. As the number of lines in 
the cluster $N$ grows, the bias velocity $v$ is reduced. Consequently the rate 
${\dot N}$ decreases gradually with time and finally at some critical value of the 
bias $v = v_{\rm cn}$ vortex formation stops altogether. The measurement  
illustrates that in constant conditions the vortex formation rate ${\dot N}$ is 
indeed controlled by the bias velocity $v$ and not by the velocity of the normal 
component $v_{\rm n}(R) = \Omega R$, which is constant during this entire 
measurement. The dashed curve through the data is fully specified by the two 
parameters, the rate parameter $\gamma$, which is a constant for a given measuring 
setup with fixed neutron flux, and the threshold velocity $v_{\rm cn}$, which 
contains the dependence on the sample variables (temperature, pressure, magnetic 
field). The large excursions of the measured data from the dashed curve illustrate 
the stochastic nature of the accumulation process. 

\begin{figure}[tp] 
\centerline{\includegraphics[width=0.9\columnwidth]{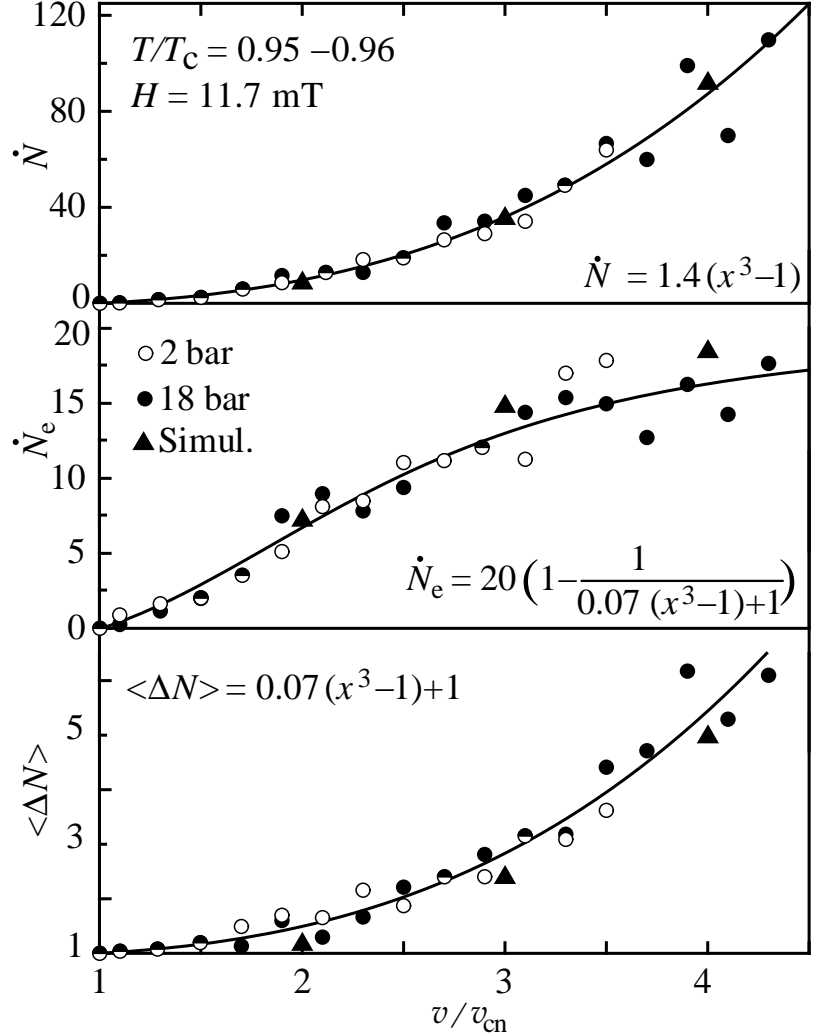}} 
\caption[BiasDependence] 
  {Different rates in neutron-induced vortex formation, 
  plotted {\it vs} normalized bias flow velocity $v/v_{\rm cn}$: {\it
    (Top)} the total number of rectilinear lines $\dot N$ formed per
  minute, {\it (middle)} the number of observed neutron absorption events
  $\dot N_{\rm e}$ per minute, and {\it (bottom)} the average number of
  lines $\langle \Delta N \rangle$ formed per observed event. All three
  rates have been counted {\it independently} from discontinuities in NMR
  absorption with single-vortex amplitude resolution and single-event time
  resolution.  The two upper plots correspond to an incident neutron flux of
  about 20 neutrons/min. The bottom plot is neutron-flux independent.  The
  solid curves are fits to the data, given by the expressions in each
  panel.  Triangles are results of numerical simulations described in
  Ref.~\protect\onlinecite{PLTP-chapter}. }
\label{BiasDependence}
\vspace{-6mm}
\end{figure}

The three panels in Fig.~\ref{BiasDependence} describe in more detail the 
comparison of the rate equation (\ref{eq20}) to measurements.\cite{Eltsov} From the 
staircase like NMR absorption, as shown in Fig.~\ref{Saturation}, one counts both 
the escape rate for vortex loops and the frequency of successful neutron capture 
events, {\it ie.}  the events which lead to vortex-line formation. This requires a 
weak thermal neutron flux so that two neutron capture events do not overlap in 
time. In the measurements the rate was adjusted by changing the distance $d$ 
between the paraffin-moderated Am-Be neutron source and the $^3$He 
sample.\cite{PLTP-chapter} The three rates in Fig.~\ref{BiasDependence} have been 
determined independently and directly by counting the number of rectilinear vortex 
lines from each measured event separately. To construct the plot, the horizontal 
axis was divided into equal bins from which all individual measurements were 
averaged to yield the evenly distributed data points displayed in the panels. The 
rates increase rapidly with the bias velocity $v$ from the critical threshold 
$v_{\rm cn}$ to $4.5\,v_{\rm cn}$. This upper limit is close to the maximum 
possible bias velocity, imposed by the spontaneous instability limit of this 
particular sample container.

%%%%%%%%%%%%%%%%%%%%%%%%%%%%%%%%%%%%%%%%%%%%%%%%%%%%%%%%% 
%%%%%%%%%%%%%%%%%%%%%%%%%%%%%%%%%%%%%%%%%%%%%%%%%%%%%%%%%%%

The top panel in Fig.~\ref{BiasDependence} demonstrates the cubic dependence of the 
vortex formation rate $\dot N$ on the bias flow. A linear dependence would here 
fall outside the experimental uncertainty limits. Also note that in agreement with 
the scaling properties of the rate equation (\ref{eq20}), measurements at the two 
pressures of 2 and 18 bar fall on the same universal curve as a function of 
$v/v_{\rm cn}$. 
  
The center panel illustrates the rate of those neutron absorption events $\dot 
N_{\rm e}$ which produce at least one line and thus become observable in the NMR 
absorption measurement (i.e. produce a step of any size in Fig.~\ref{Saturation}). 
This plot shows that the rate of successful events increases with bias flow and 
appears to saturate at about 20 neutrons/min. This is the flux of neutrons absorbed 
in the sample, as also estimated from independent measurements with commercial 
monitoring devices for thermal neutrons.\cite{PLTP-chapter} The measurement thus 
involves a strong stochastic element -- close to the critical threshold at $v = 1.1 
v_{\rm cn}$ only one neutron capture event from 40 manages to produce a 
sufficiently large vortex loop for spontaneous expansion. On increasing the bias 
flow by a factor of four, almost all neutron capture events give rise to at least 
one escaping vortex loop. 

The bottom panel records the average number of vortex loops $\langle \Delta N 
\rangle$ which are extracted from each detected neutron absorption event (i.e. the 
average height of the step in Fig.~\ref{Saturation}). Considering all three panels 
of Fig.~\ref{BiasDependence}, we now realize that the rapid increase of $\dot N$ as 
a function of $v/v_{\rm cn}$ arises from the increase in both the event rate $\dot 
N_{\rm e}$ and the number of lines produced per event $\langle \Delta N \rangle$. 
This conclusion fits with the KZ predictions, while for the surface instability one 
would expect $\dot N_{\rm e}(v)$ to resemble a step function and $\langle \Delta N 
\rangle$ to increase linearly with $v$.

\begin{figure}[tp]
\centerline{\includegraphics[width=0.7\columnwidth]{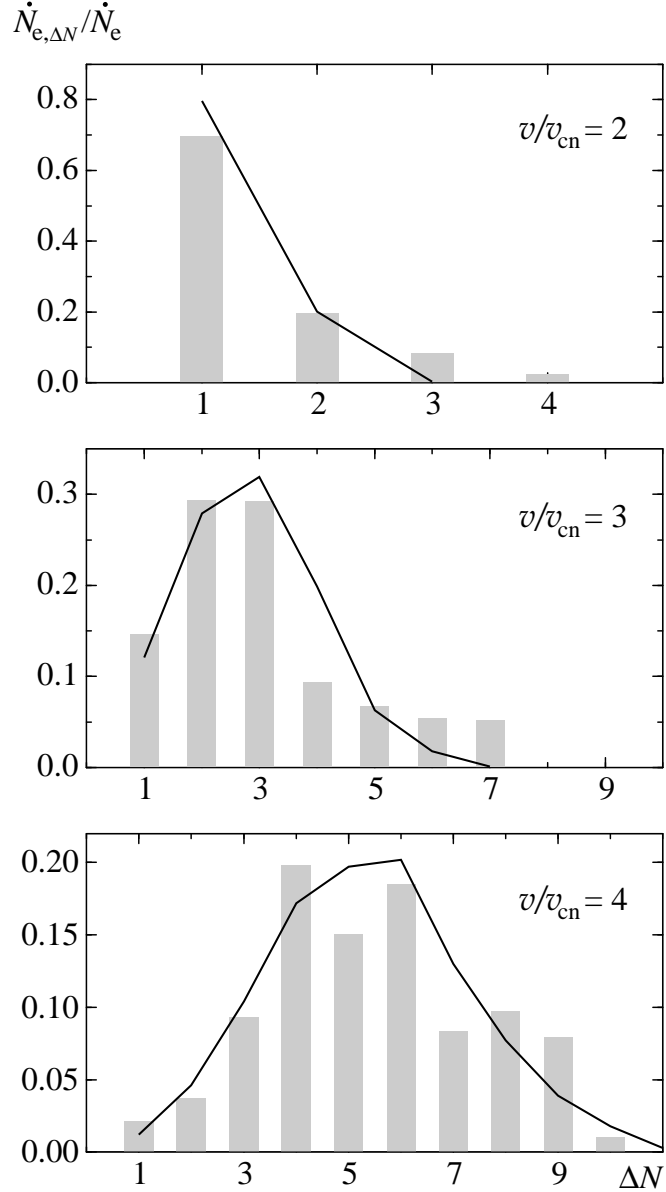}}
\caption[VorRingCount] { Distribution of neutron-induced vortex formation
  events, as a function of the number of rectilinear vortex 
  lines $\Delta N$ produced in a single event at different values of the 
  bias flow $v$ (normalized to the
  critical velocity $v_{\rm cn}$). The rate of events $\dot N_{{\rm
  e,}\Delta N}$ producing the specific number of lines $\Delta N$, normalized to the total event rate
  $\dot N_{\rm e}$, is shown. The vertical bars represent the experimental data (at
  $P = 2\,$bar, $T = 0.96\,T_{\rm c}$, and $H = 11.7\,$mT). The lines are from the 
  simulation calculations described in Ref.~\onlinecite{PLTP-chapter}.}. 
\label{VorRingCount} 
\end{figure}

The most detailed information from the rate measurements is the dispersion into 
events in which a given number of rectilinear lines $\Delta N$ is 
formed.\cite{Eltsov} Fig.~\ref{VorRingCount} displays the distribution of the 
observed events as a function of $\Delta N$, at different values of the bias flow. 
This result is the clearest demonstration for the stochastic nature of the vortex 
formation process: The width of each distribution is comparable to its average 
value $\langle \Delta N \rangle$. Such a distribution can hardly be expected to 
result from the deterministic surface instability. However, as shown in the figure, 
the distributions can be reproduced without any fitting parameters with numerical 
simulations of the evolution of a vortex network in the bias flow, if the network 
is completely random and has the of proper initial spatial 
extension.\cite{PLTP-chapter}

%%%%%%%%%%%%%%%%%%%%%%%%%%%%%%%%%%%%%%%%%%%%%%%%%%%%%%%%%%%%%
%%%%%%%%%%%%%%%%%%%%%%%%%%%%%%%%%%%%%%%%%%%%%%%%%%%%%%%%%%%%%

The superflow instability at the periphery of the neutron bubble should not be 
sensitive to the processes in the hot interior. Therefore another link to the KZ 
model would be the identification of any signatures from the primordial disorder in 
the order parameter distribution which was originally precipitated during the 
quench cooling. Several observations point to such evidence. One of these is the 
fact\cite{SpinMass} that the spin-mass vortex is formed in the neutron absorption 
process at high bias flow, when $v/v_{\rm cn} \gtrsim 2$. The spin-mass vortex is a 
combined defect. It consists of a usual mass-flow vortex (resulting from broken 
$U(1)$ symmetry) embedded within a domain-wall-like soliton (resulting from broken 
relative $SO(3)$ symmetry in the rotation of the spin and orbital coordinate 
systems with respect to each other). It is difficult to explain the formation of 
this composite in the context of a flow instability, since mass flow does not 
interact directly with the spin of the order parameter.  In contrast, fluctuations 
in a non-equilibrium transition could be expected to produce different types of 
inhomogeneity in a multi-component order parameter distribution.

The same argument holds for the competition between A and B phase
order-parameter components as a function of pressure or magnetic field.
Measurements show\cite{Eltsov} that there is an abrupt increase in $v_{\rm
  cn}$ and a corresponding reduction in the vortex formation rate $\dot N(v)$, when
the pressure is increased above the polycritical point in the superfluid $^3$He 
phase diagram. The increase in $v_{\rm cn}$ is larger than one would expect from 
the decrease of the bubble size with pressure, according to Eq.~(\ref{e.2}). In 
this pressure regime a sliver of stable A phase exists below $T_{\rm c}$ in the 
equilibrium phase diagram. The measurement is performed at a bath temperature $T_0$ 
which corresponds to the stable B-phase regime, but in the quench cooling process 
the existence of the A-phase sliver can be expected to interfere. Thus the observed 
reduction in $\dot N$ has been interpreted to indicate that with increasing 
pressure larger regions of the primordial order parameter distribution after the 
rapid cool-down of the neutron bubble are occupied by A-phase seeds which do not 
contribute to the formation of the B-phase vortex network.\cite{KH} In this way the 
available volume for the B-phase network is reduced. The preference of order 
parameter fluctuations to promote A-phase components with increasing pressure has 
been discussed by Bunkov {\it et al.} in the context of non-equilibrium 
transitions.\cite{Yuriy} Similarly with increasing magnetic field the fluctuations 
are likely to be biased more and more towards the A-phase regime of the order 
parameter space. 

The presence of the A-phase connected pressure and magnetic field dependences have 
been established in measurements,\cite{Eltsov} but more details of these features 
should be worked out.  These phenomena are believed to be related to the 
observation\cite{Osheroff} that in supercooled A phase a transition to B phase can 
be triggered by a neutron absorption event, while no transitions are observed in 
the absence of the neutron flux. The current view holds that in supercooled A phase 
homogeneous B-phase nucleation is impossible,\cite{Leggett} but that extrinsic 
effects, of which only ionizing radiation has been clearly identified so far, 
become responsible for the A$\rightarrow$B transitions which in practice are always 
eventually seen when the temperature is reduced. 

\section{VORTEX FORMATION BELOW {\boldmath$0.6\,T_{\bf c}$}}
\label{LowT-measurement}

The measurements of neutron-induced vortex formation in rotating $^3$He-B were 
formerly not extended below $0.80\,T_{\rm c}$ because of resolution difficulties: 
In the B phase, the susceptibility drops with decreasing temperature, while the 
width of the NMR spectrum increases. Thus the change of NMR absorption per one 
rectilinear vortex line quickly diminishes and single-vortex resolution is lost. 
Recently a new phenomenon was discovered:\cite{Turbulence} If a few seed vortex 
loops are injected into rotating vortex-free superflow in the B phase at 
temperatures $< 0.6\,T_{\rm c}$, they become easily unstable and form a turbulent 
vortex tangle. In the turbulent state the number of vortices rapidly multiplies and 
the final state is an equilibrium array of rectilinear lines. The change from 
vortex-free to equilibrium rotation is easy to detect with NMR at all temperatures! 
Neutron irradiation provides a convenient, almost temperature-independent way to 
inject locally into the flow vortex loops. Their number can be tuned in the range 
1--5 by choosing the bias velocity (see Fig.~\ref{VorRingCount}). Thus it became  
instructive to study the transition to turbulence when the seed loops are provided 
by a neutron absorption event. Below we examine this transition and present 
therefore results on both regular and turbulent vortex formation in neutron 
irradiation at temperatures $0.4 < T/T_{\rm c} < 0.6$.

\subsection{Experimental techniques} \label{experimental}

%%%%%%%%%%%%%%%%%%%%%%%%%%%%%%%%%%%%%%%%%%%%%%%%%%%%%%%%%
\begin{figure}[tp]
\centerline{\includegraphics[width=0.64\linewidth]{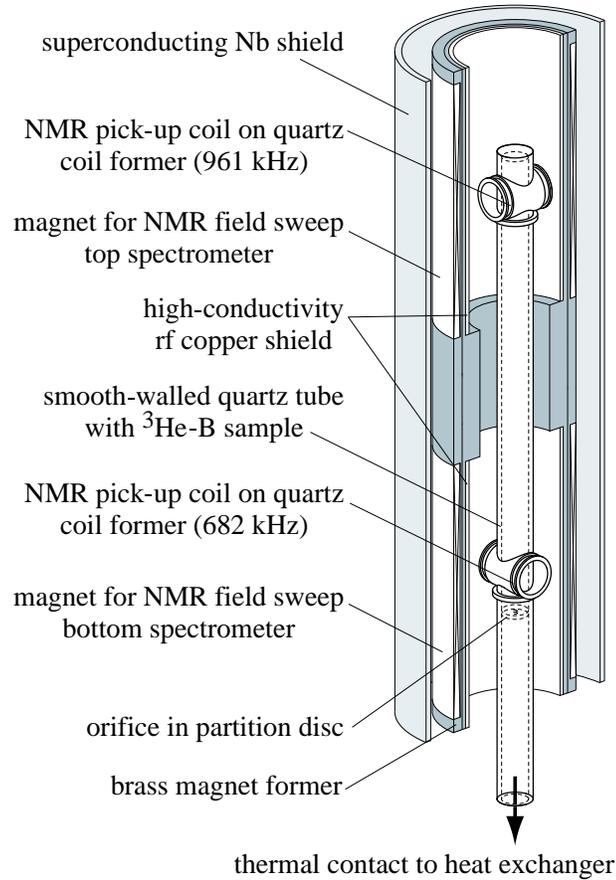}}
\caption{$^3$He sample with NMR measuring setup.\cite{Rob} The sample is
  contained in a fused quartz glass tube which has a diameter of 6\,mm and
  length 110\,mm. This space is separated from the rest of the liquid
  $^3$He volume with a partition disc.  In the disc an orifice of 0.75\,mm
  diameter provides the thermal contact to the liquid column which connects
  to the sintered heat exchanger on the nuclear cooling stage. Two
  superconducting solenoidal coil systems with end-compensation sections
  produce two independent homogeneous field regions with axially oriented
  magnetic fields. An exterior niobium cylinder provides shielding from
  external fields and additional homogenization of the NMR fields. The NMR
  magnets and the Nb shield are thermally connected to the mixing chamber
  of the pre-cooling dilution refrigerator and have no solid connection to
  the sample container in the center. The two split-half detection coils
  are fixed directly on the sample container, are wound from
  superconducting copper-nickel clad multi-filamentary wire of 0.05\,mm
  diameter, and have two layers of windings ($2\times26$ turns) in each
  half. To minimize rf losses high conductivity copper shields are
  installed inside the bores of the magnets.} \label{ExpSetUp}
\end{figure}
%%%%%%%%%%%%%%%%%%%%%%%%%%%%%%%%%%%%%%%%%%%%%%%%%%%%%%%%

Our measurements are performed in the setup\cite{Rob} shown in 
Fig.~\ref{ExpSetUp}.  For the study of turbulence it proved fortunate that this 
arrangement includes two NMR detection coils at both ends of the long sample 
cylinder. A second bonus was the relatively high critical velocity of the container 
so that vortex-free superflow could be maintained to high rotation velocities. The 
sample cylinder was prepared from quartz glass by fusing the two flat end plates 
with two sections of tubing under an acetylene flame.  Subsequently the finished 
glass structure was annealed in an oven, carefully cleaned with solvents, and 
mildly etched with dilute HF.

Experience shows that on an average sample containers with such surfaces display 
relatively high critical velocities and low trapping of remanent vortices. However, 
the exact critical velocity of a sample container is a property which is not in 
good control: Presumably one bad spot on the cylindrical surface or even a loose 
dirt particle may spoil the result. A number of similar containers have been used 
with sizeable variation in their critical properties. In one case it was observed 
that the critical velocity dropped by 50\,\% in the middle of the experiment, after 
reducing the pressure from 34 to 29\,bar, presumably because of a dislodged dirt 
particle. The latest sample tube, with which the data in this report were 
collected, had a critical velocity which was above 3.5\,rad/s (which is the safe 
upper rotation limit of our cryostat) below $0.8\,T_{\rm c}$ at 29.0\,bar.  

The lower section of the $^3$He volume below the orifice is directly connected with 
the sintered heat exchanger. This section is flooded with remanent vortices from 
the porous sinter already at low rotation. Generally we find that vortices leak 
more and more through the orifice into the sample volume on cooling below 
$0.6\,T_{\rm c}$.  Surprisingly the sample tube, with which the present 
measurements were performed, was immune to this problem in spite of the fact that 
the orifice had a relatively large diameter of 0.75\,mm.  Thus in this sample 
container vortex formation in neutron irradiation can be studied up to 3.5\,rad/s 
above $0.40\,T_{\rm c}$ if care is exercised to avoid weakly trapped remanent 
vortices.\cite{KelvinWaveInstability} It was noted that at low temperatures 
successive accelerations to rotation have to be separated by extensive waiting 
periods at stand still, to allow remanent vortices to slowly annihilate. Just below 
$0.60\,T_{\rm c}$ this waiting time proved to be around 5\,min, but at 
$0.40\,T_{\rm c}$ it was found to be of order 30\,min.  Otherwise vortex-free 
rotation at velocities above 1\,rad/s was not possible to achieve.

The sample tube is filled and pressurized with liquid $^3$He while the nuclear 
cooling stage is maintained at temperatures below 0.15\,K. In this way a new sample 
can be cooled to the lowest temperatures in one week, in spite of the large thermal 
resistance of the long liquid $^3$He column and the orifice. In the superfluid 
temperature range the thermal response of the sample cylinder is fast, the thermal 
gradient along the column is small, and the sample was cooled to $0.3\,T_{\rm c}$ 
at pressures above 10\,bar.

\begin{figure}[tp]
\begin{center}
\leavevmode
\includegraphics[width=1.0\linewidth]{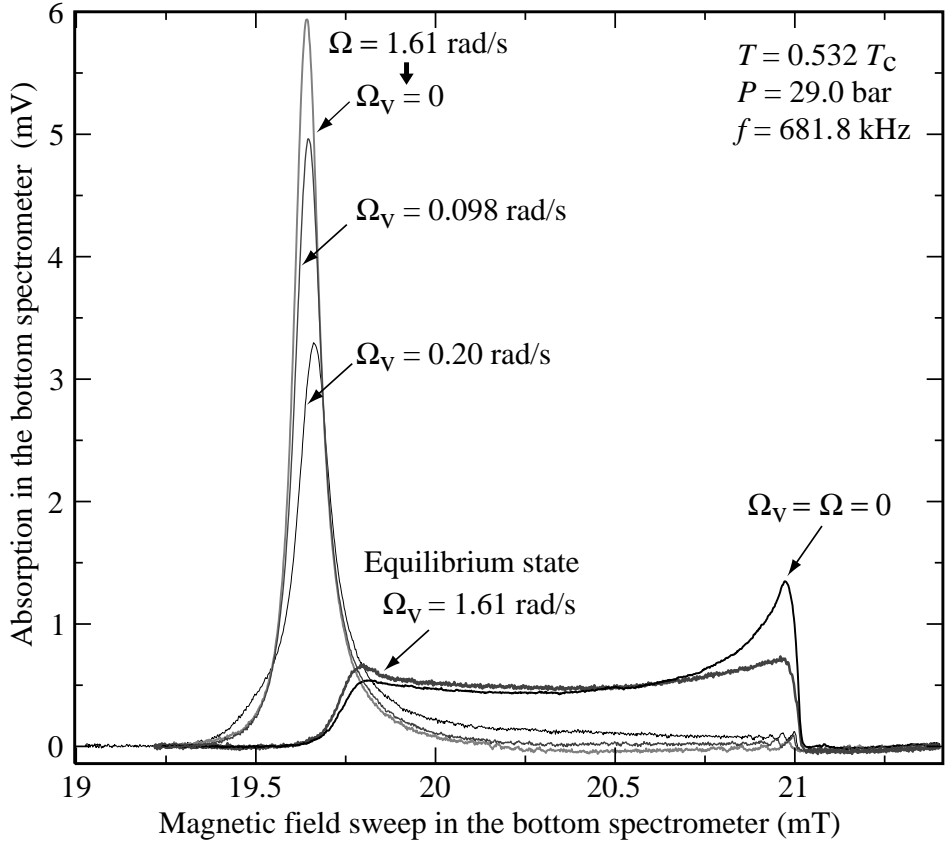}
\caption{NMR absorption spectra of $^3$He-B in rotation.
  With large vortex-free superflow the shape of the spectrum is a sensitive
  function of the number of vortex lines $N$. Here $N$ is characterized
  by the rotation velocity $\Omega_{\rm v}(N)$ at which a given number of lines
  $N$ is in the equilibrium state. The different spectra have been measured
  with the RF excitation at constant frequency $f$, using a linear sweep of the
  axially oriented polarization field $H$. The spectra have been recorded at
  constant temperature and thus all have the same integrated total
  absorption. The sharp absorption maximum at low field is called the {\it
  counterflow peak} (CF). Its shift from the Larmor field (at 21.02\,mT)
  is used for temperature measurement. When a central cluster of
  rectilinear vortex lines is formed, the height of the CF peak is reduced.
  In the {\it equilibrium vortex state} $(\Omega = \Omega_{\rm v})$, 
  where the number of vortex lines reaches
  its maximum, the spectrum looks very different: it has appreciable
  absorption at high fields and borders prominently to the Larmor edge.
  This spectrum is more similar to that of the {\it non-rotating state}
  $(\Omega = 0)$. As shown in Fig.~\protect\ref{ClusterCalibration}, when
  the vortex number is small, $\Omega_{\rm v} \ll \Omega$, the reduction in
  the CF peak height can be conveniently calibrated to give $\Omega_{\rm
  v}$ and thus $N$.  }
\label{NMR-Spectra}
\end{center}
\vspace{-6mm}
\end{figure}

The experimental setup is equipped with two independent continuous-wave NMR
spectrometers. Each spectrometer includes a split-half excitation/detection coil. 
The two coils are installed at both ends of the sample cylinder. Each coil is part 
of a high-Q tank circuit. For the upper coil the resonance frequency of the tank 
circuit is 961.2\,kHz and the Q-value 10800. The tank circuit with the lower coil 
is tuned to 681.8\,kHz and has a Q of 8800. Each spectrometer operates as a Q meter 
and is equipped with a GaAs MESFET preamplifier, which is cooled to liquid He 
temperature  and is followed by room-temperature phase-locked detection.  
Fig.~\ref{NMR-Spectra} shows a number of absorption spectra measured with the 
bottom spectrometer. The spectra illustrate the relevant NMR line shapes which one 
encounters with varying numbers of rectilinear vortex lines in the sample.  To 
obtain reproducible line shapes the sample needs to be accelerated to high rotation 
and in different rotation directions, while still at high temperatures around 
$0.8\,T_{\rm c}$. This homogenization procedure improves the order parameter 
texture by displacing solitons and other defects from the sample volume.

%%%%%%%%%%%%%%%%%%%%%%%%%%%%%%%%%%%%%%%%%%%%%%%%%%%%%%%%
%%%%%%%%%%%%%%%%%%%%%%%%%%%%%%%%%%%%%%%%%%%%%%%%%%%%%%%%

Three spectra in Fig.~\ref{NMR-Spectra} have a sharp maximum on the left which 
arises from the vortex-free bias flow. It is traditionally called the {\it 
counterflow} (CF) peak. If vortices are formed, these accumulate as rectilinear 
lines in a central vortex cluster. The superflow outside the cluster is then 
reduced and both the height and total intensity of the CF peak drop.  The reduction 
is proportional to the number of new rectilinear vortex lines if the overall change 
from the vortex-free state remains sufficiently small. In contrast, the line shape 
looks very different in the equilibrium rotating state, when the entire sample is 
filled with rectilinear lines and the large-scale vortex-free superflow is 
practically absent. As a function of vortex number $N$, the change from the 
vortex-free to the equilibrium spectrum is by no means linear at constant $\Omega$. 
In fact close to the equilibrium state ($\Omega - \Omega_{\rm v}(N)\lesssim 
0.2\,$rad/s), the spectrum is already almost insensitive to any changes in $N$. 
Here we denote with $\Omega_{\rm v}(N)$ the rotation velocity at which a given 
number of vortex lines $N$ is in equilibrium rotation. Therefore in the general 
case, to determine $N$ one has to increase rotation to some reference value where 
the sensitivity of the CF peak is restored.\cite{Parts} This requires that the 
textures are stable and reproducible and that new vortices are not formed during 
rotational acceleration.  In the neutron irradiation measurements below we 
primarily need to distinguish two cases from each other, the production of a small 
number of rectilinear vortex lines or the equilibrium number after a turbulent 
event. As seen from Fig.~\ref{NMR-Spectra}, this task can be accomplished just by 
inspection of the spectra without any calibration.

%%%%%%%%%%%%%%%%%%%%%%%%%%%%%%%%%%%%%%%%%%%%%%%%%%%%%%%%
\begin{figure}[t]
\centerline{\includegraphics[width=0.8\linewidth]{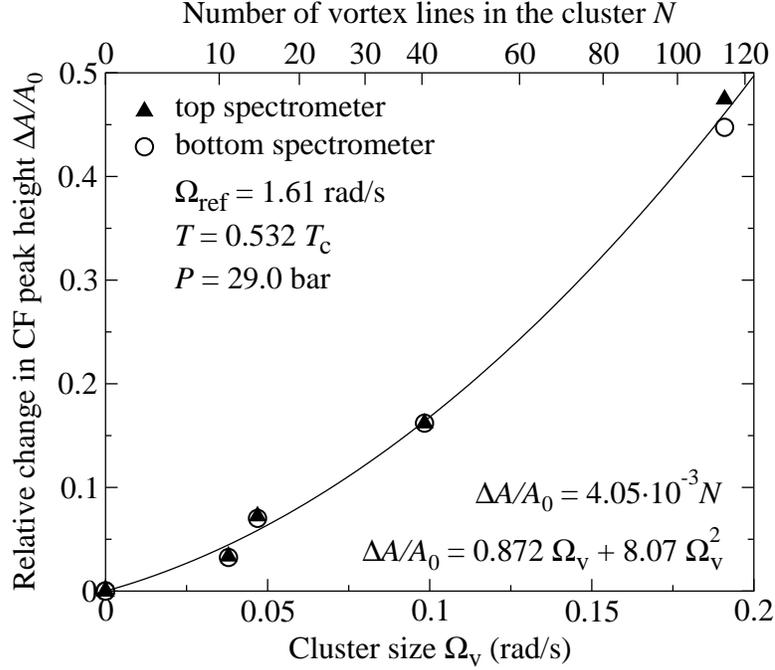}}
\caption{Calibration of CF peak height $A$ versus vortex line number $N$.
  The CF peak height in the vortex-free state, $A_0(\Omega_{\rm ref})$, is
  compared to the peak height $A(\Omega_{\rm ref},\Omega_{\rm v}(N))$,
  which is measured at the same rotation velocity $\Omega_{\rm ref}$ for a
  small vortex cluster of size $\Omega_{\rm v}$, which is prepared as
  described in the text. The quantity plotted on the vertical scale is the
  relative reduction in peak heights, $\Delta A/A_0 = [A_0(\Omega_{\rm
    ref}) - A(\Omega_{\rm ref},\Omega_{\rm v}(N))]/A_0(\Omega_{\rm ref})$,
  measured at constant conditions. The solid line is a fit, given by the expression
  in the figure. The conversion from $\Omega_{\rm v}$ (bottom axis) to
  $N$ (top axis) in the continuum picture is\cite{Ruohio} $N = N_0
  (1-d_{\rm eq}/R)^2$, where $N_0 = \pi R^2 (2 \Omega_{\rm v}/\kappa)$ and
  $d_{\rm eq} = [(\kappa/8\pi\Omega_{\rm v}) \, \ln(\kappa/2\pi\Omega_{\rm
    v} r_{\rm c}^2)]^{1/2}$. Here $r_{\rm c} \sim \xi(T,P) $ is the radius
  of the vortex core. Using this conversion it is found that $\Delta A/A_0$
  is a linear function of $N$, similar to Ref.~\onlinecite{Xu}. }
\label{ClusterCalibration}
\end{figure}
%%%%%%%%%%%%%%%%%%%%%%%%%%%%%%%%%%%%%%%%%%%%%%%%%%%%%%%%

For measurements of the vortex formation rate in non-turbulent conditions a 
calibration of the CF peak height with respect to the number of vortex lines is 
required.  Fig.~\ref{ClusterCalibration} shows an example of a such calibration. 
The measurement was performed for small vortex clusters ($N \lesssim 100)$ using 
the following procedure: A vortex cluster of given size is formed and measured in a 
four step process. 1) First the spectrum is recorded in the vortex-free state in 
the reference conditions and the CF peak height $A_0(\Omega_{\rm ref})$ is 
obtained. 2) Next a large number of vortex lines is created by irradiating with 
neutrons (with the rotation at $\Omega_{\rm ref}$ or higher).  3) The sample is 
then decelerated to the low rotation velocity $\Omega_{\rm low} \ll \Omega_{\rm 
ref}$ so that part of the vortex lines are observed to annihilate. Since our long 
cylinder is oriented along the rotation axis only within a precision of $\sim 
0.5^{\circ}$, any annihilation barrier must be negligible. This means that at 
$\Omega_{\rm low}$ the sample is in the equilibrium vortex state $\Omega_{\rm low} 
= \Omega_{\rm v}(N)$ with a known\cite{Ruohio} number of vortex lines $N$. 4) 
Finally the rotation is increased back to the reference value $\Omega_{\rm ref}$ 
and the spectrum is recorded in order to measure the CF peak height $A(\Omega_{\rm 
ref}, \Omega_{\rm v}(N))$.  In Fig.~\ref{ClusterCalibration} the reduction $\Delta 
A = A_0(\Omega_{\rm ref}) - A(\Omega_{\rm ref}, \Omega_{\rm v})$ in the CF peak 
height of the two spectra is plotted as a function of $\Omega_{\rm v}$. To reduce 
the dependence on drift and other irregularities we normalize the reduction $\Delta 
A$ to the CF peak height $A_0(\Omega_{\rm ref})$ of the vortex-free reference 
state.  In the limit $\Omega_{\rm ref} \gg \Omega_{\rm v}$, the result is a smooth 
parabola. If $\Delta A/A_0$ is plotted as a function of number of vortex lines $N$, 
calculated from $\Omega_{\rm v}$, then the dependence is linear.

Any sample with an unknown number of vortex lines, which is less than the maximum 
calibrated number, can now be measured in the same reference conditions 
($\Omega_{\rm ref}$, $T$, and $P$) and compared to this plot, to determine $N$.  In 
Fig.~\ref{VortexProductionRate}~~ a measurement of vortex formation in neutron 
irradiation is shown for which the calibration plot in 
Fig.~\ref{ClusterCalibration} was used. After each irradiation session at different 
bias flow velocity $v= \Omega R$ the rotation velocity is changed to $\Omega_{\rm 
ref}$ and the NMR absorption spectrum is recorded. From this spectrum the reduction 
in CF peak height is determined by comparing to the spectra of the vortex-free 
state which are measured regularly between neutron irradiation sessions. Such a 
calibration is less time consuming than other methods, if the temperature is kept 
stable during the measurements. Its accuracy relies on (i) the stability and 
reproducibility of the order parameter texture, (ii) the precision with which the 
definition of the cluster size is achieved (ie. the stability of rotation at 
$\Omega_{\rm v}$), and (iii) the validity of the continuum model and absence of 
annihilation barrier at small vortex numbers $(N \sim 10$\,--\,100).

\vspace*{-3mm}

\subsection{Measurement of vortex formation rate} \label{MeasureRate}

%%%%%%%%%%%%%%%%%%%%%%%%%%%%%%%%%%%%%%%%%%%%%%%%%%%%%%%%
\begin{figure}[t]
\begin{center}
\leavevmode
\includegraphics[width=0.9\linewidth]{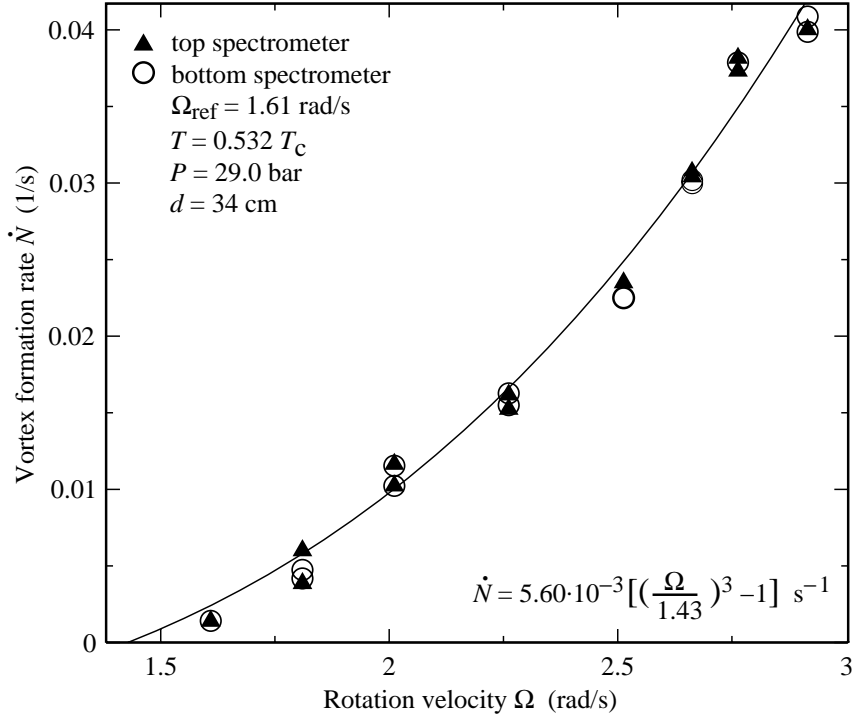}
\caption{Rate of vortex formation in neutron irradiation. The average number of
  rectilinear vortex lines created per unit time during the irradiation
  period is shown as a function of the rotation velocity $\Omega$. The data
  are fitted with the expression ${\dot N} = 0.336\,[(\Omega/1.43)^3 -
  1]\,$min$^{-1}$ (with $\Omega$ in rad/s). Depending on the rate ${\dot
    N}$, the irradiation time varies here from 30\,min to 4.5\,h, so that
  the number of accumulated vortices remains within the range of the calibration
  in Fig.~\protect\ref{ClusterCalibration}. This calibration is used to
  determine the number of vortices from the relative reduction in
  the CF peak height, $\Delta A/A_0$.  The distance\cite{PLTP-chapter} of
  the neutron source from the sample was $d = 34\,$cm. The range of the 
  bias flow is limited in these measurements between the critical velocity 
  $\Omega_{\rm cn} = 1.43\,$rad/s and the upper limit $\Omega = 3.0\,$rad/s, 
  where the turbulent events start to occur. } 
  \label{VortexProductionRate} 
  \end{center} 
\vspace{-6mm} 
\end{figure} 
%%%%%%%%%%%%%%%%%%%%%%%%%%%%%%%%%%%%%%%%%%%%%%%%%%%%%%%%

In Fig.~\ref{VortexProductionRate} the rate of vortex formation $\dot N$ in neutron 
irradiation is measured at $0.53\,T_{\rm c}$ as a function of the bias velocity $v= 
\Omega R$. The result supports the cubic rate equation (\ref{eq20}) at a lower 
temperature than has been reported previously. The critical velocity $v_{\rm
  cn} = \Omega_{\rm cn} R = 4.3\,$mm/s is consistent with the values measured
previously at pressures up to 21.5\,bar and temperatures above $0.80\,T_{\rm c}$. 
This is plausible if the critical velocity is determined only by the size of the 
neutron bubble which does not change appreciably with decreasing temperature, 
Eq.~(\ref{e.2}). However, the rate factor $\gamma$ is 36 times smaller than in 
earlier measurements above $0.8\,T_{\rm c}$ and at lower pressures. It is at 
present not known how $\gamma$ and $v_{\rm cn}$ vary at high pressures when a wide 
range of stable A phase exists between $T_{\rm c}$ and the ambient B-phase bath 
temperature $T_0$.  One might expect that the much enhanced probability of A phase 
nucleation in different parts of the neutron bubble reduces the volume occupied by 
the B phase.\cite{Yuriy} In this situation the B-phase vortex network would be 
limited to a smaller volume and perhaps fewer vortex loops would be extracted into 
the bias flow. In such a case one might expect $v_{\rm cn}$ to increase and 
$\gamma$ to be reduced. 

A second similar measurement at $P=10.2\,$bar and $T=0.57\,T_{\rm c}$ gives a 
critical velocity $v_{\rm cn} = 3.6\,$mm/s and a rate constant $\gamma$ which is 3 
times larger than the measurement in Fig.~\ref{VortexProductionRate}. At 10\,bar  
pressure A phase is not stable.  Since turbulent events at higher bias velocities 
interfere with these measurements, the available range of bias velocities 
(approximately $v_{\rm cn} < v \lesssim 2\, v_{\rm cn}$) is smaller than in the 
measurements at higher temperatures, see Fig.~\ref{BiasDependence}. Nevertheless, 
both of these low temperature measurements support the cubic form of the rate 
equation. In the absence of more extensive work as a function of pressure and 
temperature, the reasons for the reduced rate factor $\gamma$ remain so far 
unexplained.

\subsection{Superfluid turbulence in neutron irradiation} \label{Turbulence}

\begin{figure}[t] 
\centerline{\includegraphics[width=0.8\linewidth]{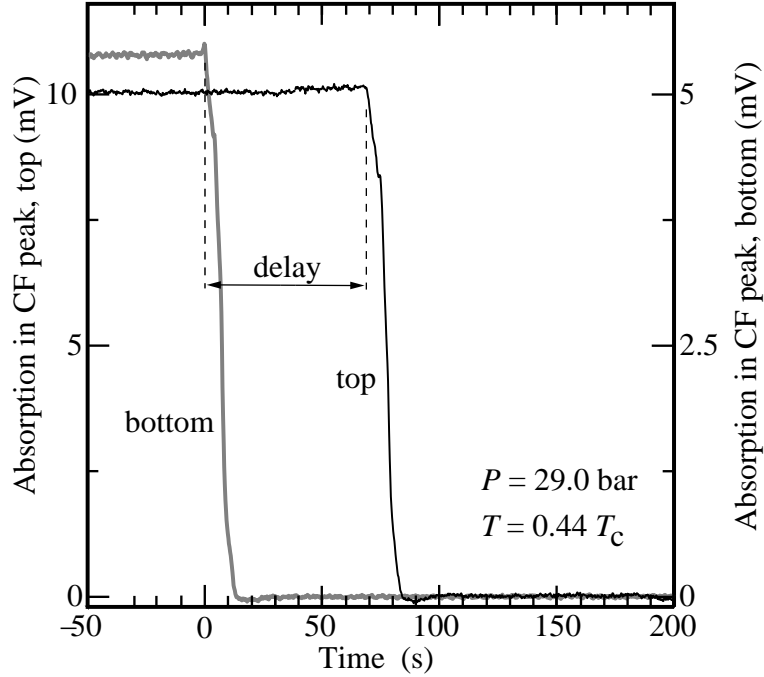}} 
\caption{NMR signatures of a vortex formation event which evolves into
  superfluid turbulence. When a vortex tangle expands into one of 
  the NMR detection coils, the height of the CF peak (Fig.~\ref{NMR-Spectra}), 
  which is monitored in this plot, rapidly drops to zero. The sequence of events is 
  schematically displayed here at $\Omega = 1.61\,$rad/s when a sudden collapse of the 
  CF peak is first observed in the bottom spectrometer (at $t=0$). After a delay 
  of 70\,s, which the turbulent front needs to travel to the lower edge of the 
  top coil 90\,mm higher along the liquid column, a similar collapse is recorded 
  by the top spetrometer. This value of delay corresponds to the situation when 
  the turbulence is first formed in the middle of the bottom
  coil. The sudden collapse of the CF peak means that the turbulent vortex tangle 
  is rapidly polarized in rotation and that the global counterflow between the normal 
  and superfluid components is thereby removed. Measurements of this type show 
  that from the initial injection site, where the injected vortex loops first start 
  to intersect in the bias flow, the turbulence expands in the rotating column by forming two 
  turbulent fronts which move at constant velocity towards the top and
  bottom ends of the sample. }
\label{TurbulenceSignal}
\end{figure}

At high rotation and temperatures below $0.60\,T_{\rm c}$ neutron irradiation 
events may become turbulent,\cite{Turbulence} similar to vortex formation from 
other sources. This means that vortex loops, which have been extracted from the 
neutron bubble and are injected into the bias flow, may start to interact, to 
produce a vortex network of large scale. This tangle then blows up and fills the 
rotating sample with the equilibrium number of vortex lines. Some NMR 
characteristics of such a neutron capture event are illustrated in 
Fig.~\ref{TurbulenceSignal}. However, the fundamental feature is that the sample is 
suddenly filled with the equilibrium number of rectilinear vortex lines, apparently 
as a result of one neutron capture event: The NMR absorption spectrum jumps (via a 
brief transitory period) from a line shape with a large CF peak to that of the 
rotating equilibrium state of totally different form (Fig.~\ref{NMR-Spectra}).

Turbulent events are observed in neutron irradiation only if (i) the rotation 
velocity exceeds $\Omega_{\rm cn}$ and (ii) the temperature is sufficiently low so 
that vortex motion is not heavily damped by mutual friction. Even then these 
processes are stochastic such that the vortex loops extracted from the neutron 
bubble only rarely achieve the proper initial conditions in which turbulent loop 
expansion starts to evolve. With decreasing temperature and increasing rotation the 
probability of turbulent events increases. This may be understood since: (i) With 
decreasing temperature mutual friction damping is reduced, Kelvin wave excitations 
grow in amplitude, more new loops are formed on the existing lines, and via 
reconnection processes the intersecting loops multiply to a turbulent cascade. (ii) 
With increasing bias velocity the number of vortex loops, which are injected into 
the bias flow, rapidly increases (Fig.~\ref{BiasDependence}, bottom) and the 
probability of their intersections increases.

%%%%%%%%%%%%%%%%%%%%%%%%%%%%%%%%%%%%%%%%%%%%%%%%%%%%%%%% 
%%%%%%%%%%%%%%%%%%%%%%%%%%%%%%%%%%%%%%%%%%%%%%%%%%%%%%%%

Fig.~\ref{TurbulenceSignal} illustrates how the height of the CF peak suddenly 
drops to zero in the NMR spectrum when a turbulent event starts to evolve.  In this 
schematic example the turbulent tangle first appears inside the bottom coil. The 
collapse of the CF peak height is the first feature in the NMR absorption spectrum 
which signals the turbulence: Its rapid decay shows that the turbulent state 
becomes polarized, to mimic on an average solid-body rotation. Simultaneously the 
NMR absorption intensity from the CF peak is transferred close to the Larmor edge 
of the spectrum where a new sharp peak rapidly grows in intensity. This peak then 
slowly decays to the line shape of the equilibrium state, see 
Fig.~\ref{NMR-Spectra}. The intensity in the Larmor region reflects how the vortex 
density evolves within the coil:\cite{Turbulence} It first rapidly overshoots to a 
value which is about twice that in equilibrium and then slowly rarefies to the 
equilibrium value.

\begin{figure}[t]
\centerline{
  \includegraphics[width=1.0\linewidth]{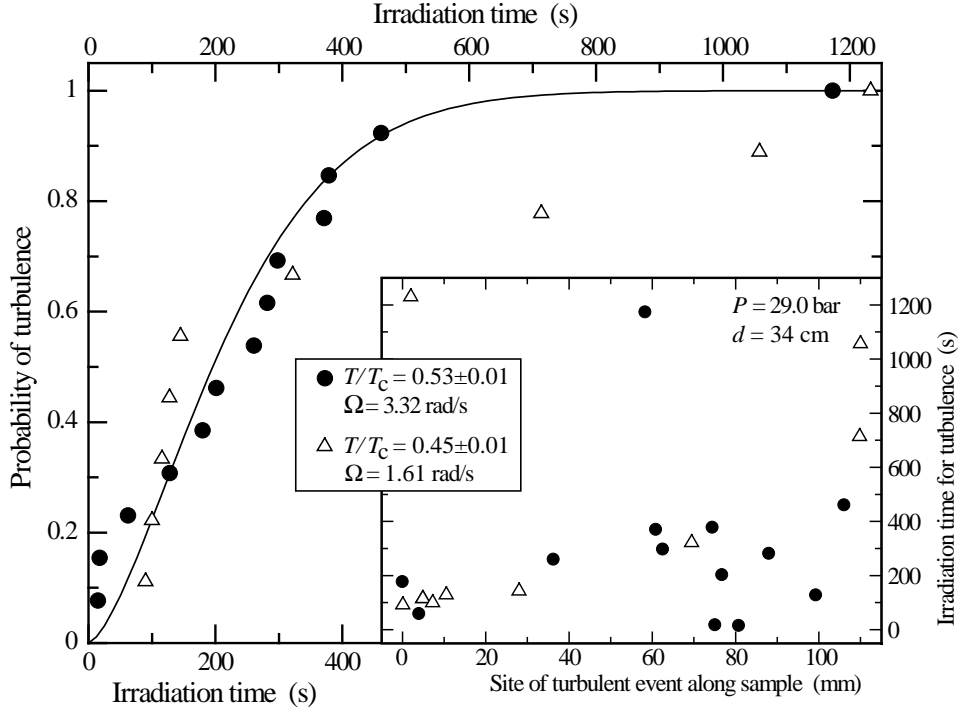}}
\caption{Turbulent vortex formation in neutron irradiation at 29.0\,bar
  pressure. The sample is irradiated in constant conditions until a
  turbulent vortex expansion event takes place. The irradiation time
  required to achieve the turbulent event is measured. Results from
  measurements at two different constant conditions are shown in this plot
  in the form of cumulative probability distributions. For comparison, the
  continuous curve represents a distribution function of the Weibull
  extreme-value form:\cite{Schoepe} $P(t) = 1-
  \exp[(-t/250\,\textrm{s})^{1.5}]$. There were no cases among the two sets
  of measurements where a turbulent event would not have been observed
  after an irradiation period of 20\,min. The {\it inset} shows
  the axial location of the initial turbulent seed, the site of the neutron
  capture, measured from the orifice upward with a technique explained in
  Fig.~\ref{TurbulenceSignal}. The corresponding irradiation time needed to
  achieve the event is shown on the vertical axis. As expected, the sites
  are randomly distributed along the sample.}
\label{NeutronTurbulence}
\end{figure}

The decaying CF signals from the two coils do not overlap in 
Fig.~\ref{TurbulenceSignal}.  Even the more slowly relaxing overshoots in the 
Larmor region do not overlap in time when $T<0.5 T_{\rm c}$ and turbulence is 
initiated at one end of the sample tube. This means that the turbulence propagates 
along the liquid column as a stratified layer: By the time the NMR absorption in 
the top coil gives the first indication of the approaching turbulent front, the 
bottom coil has already settled into its stable equilibrium line shape. Thus the 
turbulent front is formed by a relatively thin layer of disordered tangle, in which 
the polarization of the circulation reaches its final equilibrium value, before the 
vortex density and the configuration of the vortices has stabilized. The front 
moves along the column with a fixed velocity which was measured in 
Ref.~\onlinecite{FlightTime}. This velocity has the same value as that of a single 
short section of  vortex filament moving along the cylinder wall in the initial 
vortex-free bias flow: $v_{\rm z} = \alpha \Omega R$.  Here $\alpha$ is the 
dissipative mutual friction coefficient which was measured in 
Refs.~\onlinecite{FlightTime}--\onlinecite{Bevan}.  This means that from the delay 
between the signals of the two detector coils (as marked in 
Fig.~\ref{TurbulenceSignal}), we may calculate the axial location $z$ where the 
turbulent event started. This location has been plotted for the data in 
Fig.~\ref{NeutronTurbulence} in the inset of this figure.

%%%%%%%%%%%%%%%%%%%%%%%%%%%%%%%%%%%%%%%%%%%%%%%%%%%%%%%%
%%%%%%%%%%%%%%%%%%%%%%%%%%%%%%%%%%%%%%%%%%%%%%%%%%%%%%%%

The probability of a neutron capture to trigger a turbulent process is studied in 
Fig.~\ref{NeutronTurbulence}.  Two measurements are shown of the cumulative 
probability distribution of the irradiation time needed to achieve a turbulent 
event. In these two examples at different temperatures and rotation velocities the 
sample is irradiated at constant conditions until the CF peak is observed to 
collapse suddenly. The irradiation time is plotted on the horizontal axis. On the 
vertical axis the number of turbulent events observed within this time is shown, 
normalized to the total number of events: 13 events in one case (at $0.53\,T_{\rm 
c}$) and 9 in the second (at $0.45\,T_{\rm c}$). As seen from the plot, the number 
of events is insufficient to produce smooth probability distributions, but the 
irradiation times are observed to be distributed over the same range in the two 
cases, {\it ie.}  their distributions have similar average values and widths. This 
in spite of the fact that the measurements at $0.45\,T_{\rm c}$ were performed at 
half the rotation of those at $0.53\,T_{\rm c}$. At high temperatures a reduction 
by two in velocity results in a significant decrease in the yield of vortex lines 
from one neutron absorption event, Fig.~\ref{BiasDependence}. The fact that the two 
distributions in Fig.~\ref{NeutronTurbulence} do not differ significantly indicates 
that with decreasing temperature the transition to turbulence becomes more probable 
and less sensitive to the initial configuration of the injected loops.

Let us consider the measurement at $0.53\,T_{\rm c}$ in more detail. The state of 
the sample changes during the irradiation at constant $\Omega$: (i) the initial 
state is vortex-free, (ii) during the irradiation rectilinear vortex lines are 
formed at a rate which can be extrapolated from 
Fig.~\protect\ref{VortexProductionRate}, (iii) until finally all vortex-free CF is 
completely terminated in a turbulent event. In the final step the state of the 
sample changes from one with only a small central vortex cluster to one with the 
equilibrium number of vortex lines $(N_{\rm
  eq} \sim 2600)$. Note that to observe a new turbulent event the existing 
vortices have to be annihilated, by stopping rotation. Then the vortex-free state 
can be prepared again and a new irradiation session can be started. The longest 
irradiation time is here $\sim 20\,$min. During this period the cluster grows at 
the rate ${\dot N} = 3.9\,$vortices/min, so that it contains about 80 vortices when 
the turbulent event finally starts. At this point the CF velocity at the sample 
boundary $v = v_{\rm n} - v_{\rm s} = \Omega R - \kappa N/(2\pi R)$ has been 
reduced by 2.7\,\% from the initial state. Since the mean irradiation time in the 
measured distribution is only $\sim 250\,$s, the reduction in CF velocity by 
vortices formed before a turbulent event is minor. We may thus view the result in 
Fig.~\ref{NeutronTurbulence} as representative of these particular values of 
rotation and temperature.

One may wonder whether a turbulent process results from a single neutron capture 
event or from the coincidence of two or more events. In the latter case the 
simultaneous events need to be sufficiently close not only in time, but also in 
space, so that the expanding loops, which are already extracted from the two 
neutron bubbles, have a possibility to intersect. (The probability of the bubbles 
themselves to intersect is very small.) The intersection of the loops expanding 
from the two random positions along the height of the cylindrical sample is more 
likely to occur closer to the middle than at the top or bottom end. However, the 
events listed in the inset of Fig.~\ref{NeutronTurbulence} occur randomly along the 
sample. In all cases the NMR signatures from the turbulent events are similar, 
there is no prominent variation in their appearance depending on where the event 
starts.  Additionally, at lower temperatures, the turbulent events occur close 
above $\Omega_{\rm cn}$.  Here successful neutron absorption events, which lead to 
the extraction of loops to the bulk, are rare but, nevertheless, turbulent events 
become more probable. From this we presume that a single neutron capture event in 
suitable conditions must be able to start a turbulent expansion event. For a more 
careful proof of this point the measurements should be repeated as a function of 
the neutron flux.

\section{SUMMARY}

Neutron irradiation of vortex-free superflow has become a practical means of 
creating quantized vortices in $^3$He-B. Detailed knowledge exists on many of the 
experimental features, although more measurements are needed at intermediate 
temperatures and pressures, to explain the vortex formation rate as a function of 
the bias flow velocity in all situations. At low temperatures neutron irradiation 
is a useful method for the localized injection of vortex loops into vortex-free 
superflow, to study superfluid turbulence.

As to the explanation for the neutron-induced vortex formation, the measurements 
undoubtedly support a volume effect: The vortex loops escaping into the bulk 
superflow originate from a random vortex network which forms in the interior of the 
neutron bubble when it rapidly cools through the superfluid transition. The 
validity of this Kibble-Zurek mechanism of vortex formation is also confirmed by 
current numerical simulations of the time-dependent Ginzburg-Landau type. However, 
these calculations identify the superflow instability at the surface of the warm 
neutron bubble as the source for those vortex loops which manage to escape into the 
bulk superflow. To examine the interplay of these two mechanisms, it would now be 
instructive to perform quench-cooling experiments as a function of the cooling 
rate. Rapid localized overheating can be achieved with laser pulsing, as has been 
demonstrated by H. Alles et al.\cite{Alles} Such measurements would probe cooling 
rates which are slower than after a neutron absorption event, but might reveal, for 
instance, a change over from the volume to the surface mechanism with decreasing 
cooling rate. 

\section{POSTSCRIPT}

While we were running our experiments with the rotating refrigerator, the retired 
director of our laboratory, Olli Lounasmaa, used to stroll late in the evenings to 
the control table of our rotating cryostat and test his ideas on us: "The two most 
outstanding challenges to research are the Universe and the Human Brain," he said. 
"In two hundred years from now our successors will not be interested in studying 
superfluid helium - all that you need to know about it has already become common 
knowledge. But they will still try to understand the secrets of the Universe and 
they will still continue wondering whether their Brain will suffice to grasp all 
the essence of the Cosmos."

Nevertheless, Olli was impressed by the bridge which was created by superfluid 
helium measurements and their interpretation to current ideas borrowed from 
cosmology - he supported such physical generalization wholeheartedly. After all, it 
was him who originally had the courage in 1976 to apply for funding to construct a 
rotating cryostat for the sub-mK temperature range. His legacy to us was to set an 
example of interminable energy to tackle new research challenges. Right now this is 
needed while one wonders whether superfluid turbulence in $^3$He-B can be explained 
based on the existing knowledge and expertise. This is just one of the many 
important problems to be solved before helium superfluids can be declared to be 
understood.

{\bf Acknowledgements:} We thank Carlo Barenghi, Rob Blaauwgeers, Nikolai Kopnin, 
Ladislav Skrbek, Makoto Tsubota, Joe Vinen, and Grigory Volovik for valuable 
discussions and collaboration during our recent work on superfluid turbulence. This 
collaboration has enjoyed the extra resources made available by the EU-IHP ULTI 3 
visitor program and the two ESF research programs COSLAB and VORTEX.

%\vspace{-5mm}

\end{document}